\documentclass[aps,preprint,groupedaddress,showpacs,pre]{revtex4}

\usepackage{graphicx}
\usepackage{dcolumn}
\usepackage{bm}
\usepackage{amssymb}

\begin{document}


\title{Characterization of the Drag Force in an Air-Moderated Granular Bed}


\author{Theodore A. Brzinski III}
\email{brzinski@physics.upenn.edu}
\author{Douglas J. Durian}
\affiliation{Department of Physics and Astronomy, University of Pennsylvania, Philadelphia, Pennsylvania 19104-6396, USA}


\date{\today}

\begin{abstract}
We measure the torque acting on a rod rotated perpendicular to its axis in a granular bed, through which an upflow of gas is utilized to tune the hydrostatic loading between grains. At low rotation rates the torque is independent of speed, but scales quadratically with rod-length and linearly with depth; the proportionality approaches zero linearly as the upflow of gas is increased towards a critical value above which the grains are fluidized. At high rotation rates the torque exhibits quadratic rate-dependence and scales as the rod's length to the 4th power. The torque has no dependence on either depth or airflow at these higher rates. A model used to describe the stopping force experienced by a projectile impacting a granular bed can be shown to predict these behaviors for our system's geometry, indicating that the same mechanics dictate both steady-state and transient drag forces in granular systems, regardless of geometry or material properties of the grains.
\end{abstract}


\maketitle

\section{Introduction}
Granular systems are ubiquitous in both nature and industry. While such systems are comprised of many mechanically simple particles, their bulk properties tend to be complex and nonlinear \cite{Wieghardt:1975p7003,Thompson:1991p6993,Fenistein:2003p6914,Behringer:2001p7187}. For example, a static pile of sand might behave as a solid, but if sheared or shaken strongly enough the pile will spontaneously begin to flow much like a liquid \cite{Liu:1998p7104}. This capacity to flow makes granular systems both interesting and useful, however traditional rheological models are not sufficient to fully describe granular flow. As a result, developing continuum approximations for granular flow is a long-standing goal.

An intuitive quantity to study in the context of granular flow is the drag exerted by a granular bed on a moving intruder. A general characterization of such resistive forces could be valuable in numerous applications -- such as predicting the outcome of a projectile impact into granular objects such as asteroids or planetoids \cite{Boudet:2006p7070,Walsh:2003p7129,Amato:1998p7077,Daniels:2004p7074,Zheng:2004p7302,Goldman:2008p7403} or determining the forces experienced by industrial mixing and digging equipment -- and may contribute to a more comprehensive understanding of granular matter~\cite{Uehara:2003p7073,Walsh:2004p7304,PicaCiamarra:2004p7335,Lohse:2004p7357,Hou:2005p7358}. Previous attempts to formulate such a general characterization in the context of impact dynamics have resulted in several contradictory models for the drag force~\cite{Goldman:2008p7403,Uehara:2003p7073,Walsh:2004p7304,PicaCiamarra:2004p7335,Lohse:2004p7357,Hou:2005p7358,Walsh:2003p7076}.

In a recent study of impact dynamics for a 2.54~cm diameter steel sphere, Katsuragi and Durian~\cite{Katsuragi:2007p181} demonstrated that the stopping force scales as the sum of two terms, one which grows linearly with depth, $z$, and one that grows quadratically with velocity, $v$. This functional form succesfully describes not only the results obtained in Ref.~\cite{Katsuragi:2007p181}, but also the observations reported in the numerous impact experiments which had been held to support one or another of the aforementioned contradictory drag force models.

The scaling of the two force terms observed in Ref.~\cite{Katsuragi:2007p181} suggests the following prediction for the materials dependence of drag force in the bed. The force component proportional to $v^2$ resembles the inertial drag experienced by an object moving through a fluid at high Reynolds number. Therefore, this force term ought to scale with the cross-sectional area of the intruder, $A$, and the density of the fluid - in this case the mass density, $\rho_g$, of the granular packing. The force component proportional to $z$ is reminiscent of Coulomb friction. Accordingly this force term should scale like a friction coefficient, $\mu=\tan \left( \theta _{r} \right)$ where $\theta_r$ is the angle of repose, times a hydrostatic pressure, $\rho_{g} g z$, where $g$ is acceleration due to gravity. Similar results have been obtained for the drag acting on intruders steadily driven through grains in the slow, quasistatic limit~\cite{Albert:1999p623,Albert:2000p620,Albert:2001p1481,Albert:2001p2}. By this reasoning we expect the full material dependence of the drag force to be of the form
\begin{equation}
	\label{eq:Force} 
		F=\alpha \mu \rho _{g}gAz+\beta \rho_{g}Av^{2} 
\end{equation}
In Ref.~\cite{Katsuragi:2007p181} the numerical prefactors were determined by fit to be $\alpha = 26\pm3$ and $\beta = 1.0\pm0.1$. If the materials-dependence of Eq.~(\ref{eq:Force}) is correct, these values should be independent of the details of the system, however this has yet to be tested.

While impact is a natural context in which to observe granular drag on an intruder, there are some drawbacks to using impact experiments to carefully characterize drag forces. Impacts are transient, with conditions that are inherently time dependent. Variables such as velocity and depth are entangled in an impact, adding uncertainty to the observed scaling of stopping force with these parameters. Furthermore, the drag force is generally determined by measuring instantaneous position or velocity rather than by direct measurements of stopping force, thus limiting the force resolution of experimental data. In order to address these limitations we develop a `steady-state impact' experiment by rotating a horizontal rod that is submerged in a granular bed, and measuring the torque acting on the rod once the drag has reached a steady-state. This design enables us to vary control parameters such as rate and depth in isolation, and to make measurements of drag torque directly.

Our apparatus is similar to geometries that have been utilized for the study of granular flow in the past. The most prevalent such systems are bladed mixers and vane rheometers. Bladed mixers consist of short, angled vanes rotated or pushed through a granular bed. Studies utilizing bladed mixers tend to focus on the geometry of the flows generated by the motion of the blades, emphasizing mixing efficiency rather than the explicit form of the torque acting on the blades~\cite{Remy:2009p8064,Jaworski:2001p8338,Stewart:2001p4257,Bagster:1967p1899}. Vane rheometers are devices in which a number of vertical vanes, generally spanning the entire depth of a granular bed, are rotated about a vertical axis~\cite{Barnes:2001p9297,Daniel:2008p9330,Soller:2006p8632,Poloski:2006p9438}. Vane rheometers attenuate the complications caused by wall-slip in the more rheologically conventional Couette geometry, and are primarily utilized to characterize yield stresses, thixotropic properties, and flow-profiles~\cite{Barnes:2001p9297}. In this geometry, rates and torques agree qualitatively with the model presented in Ref.~\cite{Katsuragi:2007p181}, but are considered in terms of stresses along the surface of the cylinder of granular media that is imagined to rotate as a solid along with the vanes~\cite{Daniel:2008p9330,Soller:2006p8632,Poloski:2006p9438}. In contrast, in this paper we picture the forces created by the motion of an intruder moving through a medium that is otherwise at rest.

In addition, we seek to test our interpretation of the mechanical origins of granular drag. As discussed above, contact forces in the granular bed are assumed to scale as the hydrostatic pressure independent of force chains activated by intruder motion. We modify this pressure by generating a flow of gas through the granular bed which, by Darcy's law, generates a pressure drop proportional to the gas speed through the bulk. We extend the model to account for this effect by making the substitution:
\begin{equation}
	\label{eq:pressure} 
		\rho_{g}gz\rightarrow\left\{\begin{array}{cc} \left( 1-U/U_{c} \right)\rho _{g}gz & U<U_{c} \\ \rm{unknown} & U>U_{c} \end{array}\right.
\end{equation}
where $U$ is the superficial gas speed equal to the gas flux divided by the sample cross-section, and $U_{c}$ is a critical gas speed which is sufficient to fluidize the bed. According to this substitution the only role the airflow plays is to reduce contact forces, and will thus have no effect on the inertial component of the drag. Air has been shown to play a subtle but dramatic role in the formation of convection-like patterns in shaken systems~\cite{Pak:1995} and the dynamics of granular jets generated after an impact~\cite{Royer:2005,Lohse:1234}, so such a simple dependence on gas speed is an interesting result in and of itself. For $U>U_{c}$ we expect a qualitative change in dynamics, and that the the system will cease to have a yield stress.

We verify that our experimental observations agree with this drag force model. These results reinforce the supposition that the dynamics of drag are a universal characteristic of granular systems, and that dynamics at low rates are set by static contact forces while for rapid perturbations drag is predominantly inertial, and that these dynamical regimes are universal traits of granular drag.

\section{Materials and Methods}
The experiments were performed with an Anton-Paar UDS200 rheometer. The rheometer drives the rotation of a horizontal rod, and measures the torque and corresponding rotation rate experienced by the rod. The rheometer provides us with resolution in torque as fine as $10^{-5}$ Nm, and rate resolution of $10^{-4}$ rotations per second. We are able to access a dynamical range of five orders of magnitude in both rate and torque.

\begin{figure}
	[htbp] \centering 
	\includegraphics[width=2.25in]{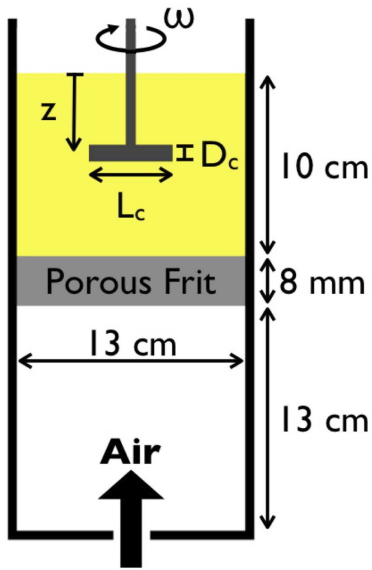}
	\caption{This is a sketch of the dish used to apply a pressure drop across the granular bed. The lower chamber, or windbox, is pressurized with air, which results in a homogenous flow of air through the porous glass frit. During an experiment the upper chamber is filled with spherical glass beads. The upflow of air through the grains generates a pressure drop which reduces hydrostatic loading. A horizontal rod of length $L_c$ and diameter $D_c$ is plunged horizontally into the glass beads to depth $z$, then rotated perpendicular to its axis.}
	\label{fig:schematic}
\end{figure}

As depicted in Fig.~\ref{fig:schematic}, the rheometer tool is replaced with a horizontal rod which rotates about its center of mass, perpendicular to its axis. The data we present were collected using three different rods with diameters(lengths) of 12.6~mm(76~mm), 6.34~mm(102~mm), and 6.34~mm(96~mm).

In place of the rheometer's standard sample plate we mount a 12.7~cm diameter dish, also depicted in Fig.~\ref{fig:schematic}. The upper portion of the dish is filled with spherical glass beads - Potters Industries with radii of $125~\mu m$ (stock number P-0060) and $180~\mu m$ (P-0080) - to a depth of 10~cm. The lower portion is a windbox, 13~cm tall. Separating the two chambers is a porous glass frit. The frit is a Robu and Schmidt P2 porous glass disc with a thickness of 8mm, and pore-sizes ranging from 50 to 100~$\mu m$. A pressure drop across the glass frit produces a homogenous upflow of gas into the granular bed. This flow of gas through the sample has been shown to produce a uniform pressure gradient which opposes gravity\cite{Ojha:2000p94}, and provides a means to tune the magnitude of static contact forces in the bulk. We use air to pressurize the windbox, and the air is passed through a pressure regulator, desiccant air-dryer, and adjustable airflow meter before entering the windbox. The pressures in the windbox are sufficient to generate superficial gas speeds of 0-44~mm/s, well in excess of the critical fluidization airspeeds. The critical airspeed for the onset of fluidization was determined to be $U_{c}=13.7$~mm/s for 125~$\mu$m grains and $U_{c}=22.5$~mm/s for 180~$\mu$m grains.

To prepare a granular bed we begin by fluidizing the grains with dried air before beginning the experiment in order to mitigate any ambient humidity. We then set the gas speed through the sample and the depth of the rod to values for which we wish to conduct a measurement. This apparatus enables us to make two different types of measurement: static measurements in which we slowly increase the torque from zero until reaching the yield torque $\tau_y$ at which the rod begins to rotate, and dynamic measurements in which we set the rod to rotate at a series of rates and record the mean torque required to maintain that rate. When we run our experiment at a constant torque, measurements of the instantaneous rotation rate exhibit fluctuations of order 10\%, which are largest at low rotation rates, and vanish at high rotation rates. Since granular systems don't exhibit conventional rheological flows, nor does our apparatus have a conventional rheometric geometry, we present our results in terms of torques and rotation rates rather than stresses and strain rates.

\section{Data}
\begin{figure}
	[htbp] \centering 
	\includegraphics[width=3.5in]{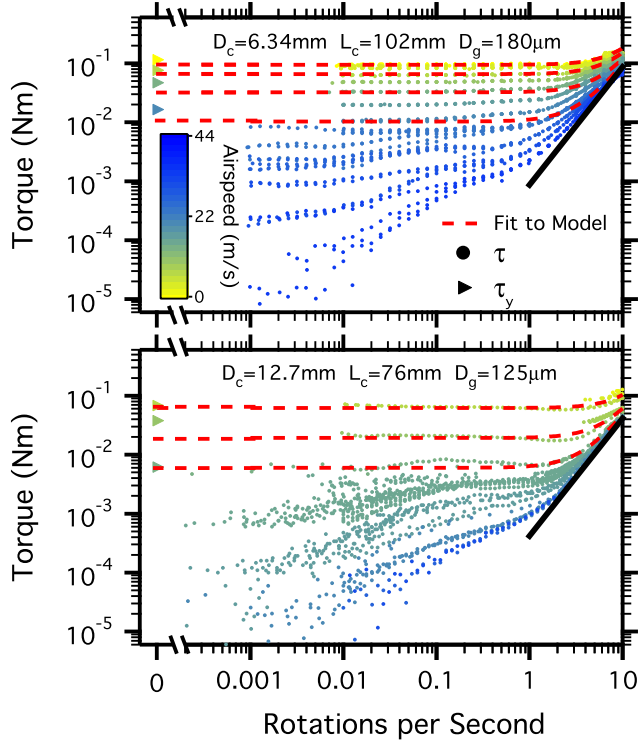}
	\caption{Torque plotted against rotation rate for gas speeds ranging from 0 to 44~mm/s. The top[bottom] plot is data from experiments conducted with 180[125]~$\mu$m grains, for which the fluidization (boiling) gas speed is around 22.5[13.7]~mm/s at a depth of 50~mm. The dashed red lines are fits to Eq.~(\protect\ref{eq:Torque}), and are all constrained to have the same value of $\beta$, corresponding to the solid black lines. The zero-rate valued triangular points denote the yield torques measured under the same conditions as the dynamics data for which fits are shown.} \label{fig:Air_fit}
\end{figure}

Example data for torque vs rotation rate are presented in Fig.~\ref{fig:Air_fit}, with upper and lower plots being for different grain and rod sizes, and where each curve corresponds to different airspeeds as labeled. At low airspeeds, and for low rotation rates, the torque is constant, hence the behavior is quasistatic. The values of the torque in this limit, $\tau_0$, decrease systematically as the upflow of air is increased. They correspond well to the yield torques, $\tau_y$, indicated by the solid triangles at zero rotation rate. At airspeeds above the critical airspeed $U_c$, at which the granular bed begins bubbling by eye, $\tau_y$ vanishes, though a nonzero torque is measured during rotation that decreases with rotation rate. At very high rotation rates, generally above one rotation per second, the measured torque increases rapidly with rotation for all airspeeds.

\begin{figure}
	[htbp] \centering 
	\includegraphics[width=3.5in]{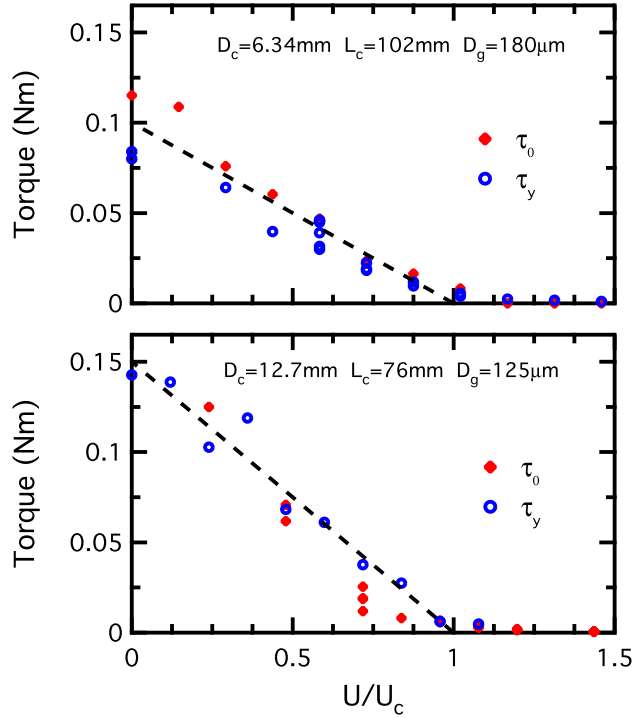} \caption{$\tau_0$ and $\tau_y$ plotted against gas speed divided by the critical fluidization gas speed for both 180~$\mu$m (top) and 125~$\mu$m (bottom) diameter grains. The lines are linear fits to the low gas speed data. $D_c$ is the rod diameter, and $L_C$ is its length.} \label{fig:Yield_Torques} 
\end{figure}

The behavior in the quasistatic regime is compared with expectation in Fig.~\ref{fig:Yield_Torques}, where both $\tau_0$ and $\tau_y$ are plotted against the normalized airspeed $U/U_c$ for the same conditions as in Fig.~\ref{fig:Air_fit}. As noted already, the values of $\tau_0$ and $\tau_y$ are in fair agreement, decreasing with $U$, and vanishing for $U>U_c$. Furthermore, the dashed lines in Fig.~\ref{fig:Yield_Torques} indicate that the quasistatic torque approaches zero at fluidization in proportion to $\left(1-U/U_{c}\right)$. This scaling is precisely that expected by Eq.~(\ref{eq:pressure}), in which grain-grain contacts are loaded hydrostatically by gravity with an offset proportional to airspeed.

To more fully test the dependence of the quasistatic torque on airspeed as well as on the details of the system, and to investigate the behavior at higher rotation rates, we have collected over one hundred sets of torque data vs rotation rate using 3 rods, 2 grainsizes, and 40 depths ranging from 2-8~cm. In all cases the results are qualitatively similar to those in Figs.~\ref{fig:Air_fit},\ref{fig:Yield_Torques}.

\section{Analysis}
To test the force model of Eqs.~(\ref{eq:Force},\ref{eq:pressure}), and ultimately deduce the values of the proportionality constants $\alpha$ and $\beta$, we compute the torque as $\int_{A}{r{\rm d}F}$ and compare with data.  In the limit that the rotating rod is long and thin, the speed and area of a differential element at distance $r$ from the rotation axis are $v=\omega r$ and ${\rm d}A=D_c {\rm d}r$, respectively, where $\omega$ is the angular rotation speed and $D_c$ is the rod diameter.  Substituting into Eq.~(\ref{eq:Force}) and carrying out the integration along the length of the rod gives the following prediction for the total torque produced by granular drag:
\begin{eqnarray}
	\label{eq:Torque}
		\tau&=&\tau_{0}+\frac{\beta}{32} \rho_{g}D_{c}L_{c}^4\omega^{2}, \\
	\label{eq:Torque0}
		\tau_{0}&=&\frac{\alpha}{4}\left(1-\frac{U}{U_c}\right) \mu \rho _{g}gD_{c}L_{c}^2z,
\end{eqnarray}
where $L_c$ is the rod length.  Fits to this form are shown as dashed curves in Fig.~\ref{fig:Air_fit}, where $\tau_0$ was adjusted independently for each data set at $U<U_c$ but where a single value of $\beta$ was enforced by simultaneous fit to all data for a given grain size and probe geometry. The inertial contribution to the torque is plotted as a solid black line.  The high quality of the fits with a single value of $\beta$ clearly demonstrate that the force model of Eq.~(\ref{eq:Force}) accurately approximates the behavior of the medium.

\begin{figure}
	[htbp] \centering 
	\includegraphics[width=3.5in]{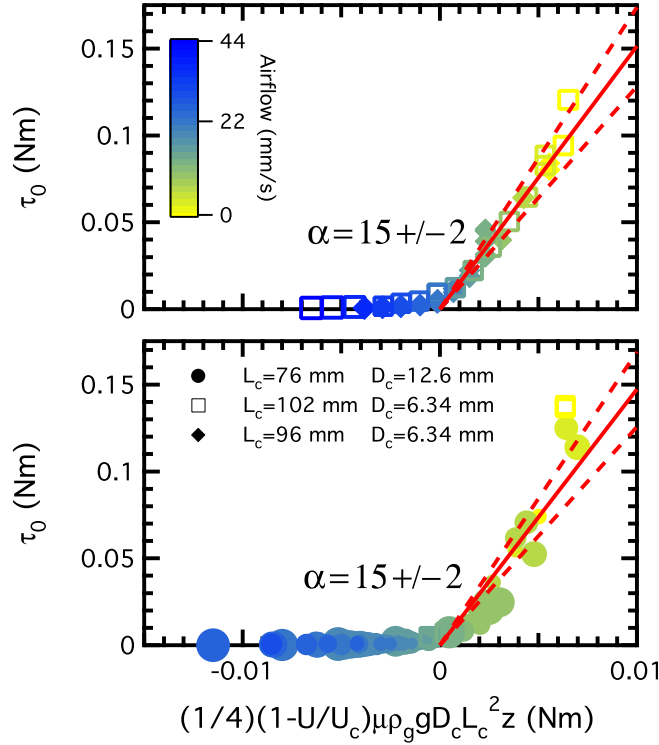} \caption{$\tau_0$ plotted against $(1/4)(1-U/U_{c})\mu\rho_{g}gD_{c}L_{c}^2z$ for both 180~$\mu$m(top) and 125~$\mu$m(bottom) grains. Coloring denotes unscaled gas speed, and symbol size denotes depth, where the largest symbols correspond to a depth of 78~mm, and the smallest correspond to depths of 20~mm.} \label{fig:alphalin} 
\end{figure}

Next we focus on experimental data for the quasistatic torque, $\tau_0$, and their comparison with the explicit materials- and geometry-dependence of Eq.~(\ref{eq:Torque0}).  For each data set, the value of $\tau_0$ is found by fitting the low rotation rate data to a constant.  The value is taken as $\tau_0=0$ for airspeeds where the torque systematically decreases as $\omega$ is reduced, as for $U>U_c$.  Results for all runs are plotted in Figs.~\ref{fig:alphalin}a,b, for the two different grain sizes, as a function of $x=(1/4) (1-U/U_c) \mu \rho_g g D_c {L_c}^2 z$.  According to the prediction of Eq.~(\ref{eq:Torque0}), both axes have units of torque and the results for $x<0$ should be zero.  But more importantly, Eq.~(\ref{eq:Torque0}) predicts that {\it all} data for $x>0$ should collapse to a line through the origin with slope $\alpha$.  Indeed this is consistent with observation, showing that Eq.~(\ref{eq:Torque0}) correctly predicts the dependence of the quasistatic torque on rod diameter, rod length, depth of rotation, and the speed of upflowing air.  Furthermore, the fits to $\tau_0 = \alpha x$ shown as lines in both Figs.~\ref{fig:alphalin}a,b give the same value $\alpha=15\pm2$ for the two different grain types, suggesting that Eq.~(\ref{eq:Torque0}) captures the complete material dependence of the quasistatic torque as well.

While the analysis of quasistatic torque thus appears in good agreement with expectation for the materials and geometry dependence, two words of caution are in order.  First, the deduced value $\alpha=15\pm2$ is somewhat smaller than the result $\alpha=26\pm3$ found in Ref.~\cite{Katsuragi:2007p181} based on the dynamics of impact of a single steel sphere into a single medium for a range of drop heights.  This could reflect actual differences due to the direction and/or geometry of intruder, horizontal rods here vs downward spheres in Ref.~\cite{Katsuragi:2007p181}; or it could reflect an uncertainty in Ref.~\cite{Katsuragi:2007p181} due to the entangling of the position- and speed-dependent force terms inherent in impact.  Second, the linear behavior of $\tau_0$ vs $x$ does not appear to hold near $x=0$, and nonzero $\tau_0$ values are even found for $x$ slightly less than zero.  We believe both effects are due to small manufacturing defects in the porous glass air diffuser, such that the speed of air is not perfectly uniform across the bottom of the sample.  A thinner region would give a higher local airspeed, and fluidize a portion of the sample at $U<U_c$; the same effect would occur for a site that persistently nucleated bubbles.  Unfortunately this is difficult to explore because the granular medium is opaque.

\begin{figure}
	[htbp] \centering 
	\includegraphics[width=3.5in]{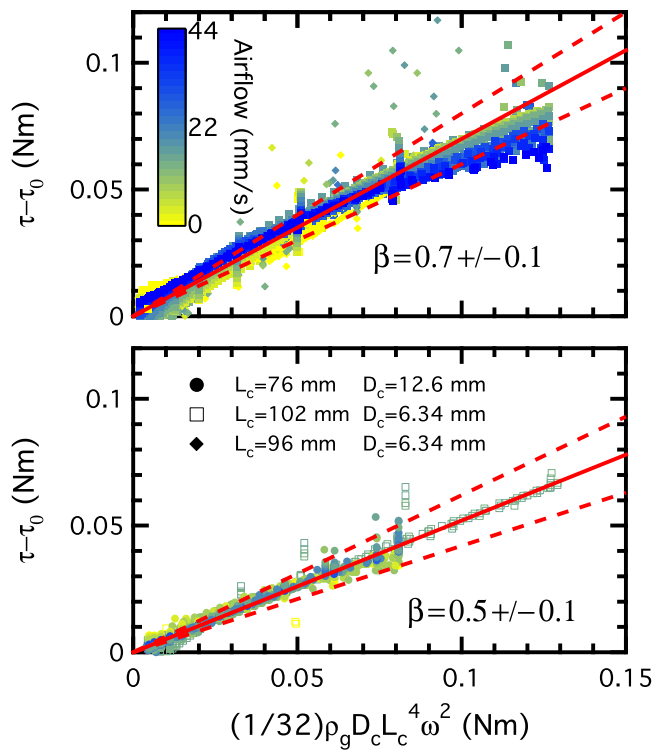} \caption{Torque plotted against $(1/32)\rho D_{c}L_{c}^4\omega^2$, and collapsed by subtracting off the quasistatic term from Eq.~(\protect\ref{eq:Torque}). The data on the top[bottom] plot is from 180[125]~$\mu$m grains, depths between 20 and 78mm, and gas speeds between 0 and $44mm/s$. For $U>U_{c}$ we take $\tau_{0}$ to be 0. The lines have slopes equal to $\beta$ and go through the origin.} \label{fig:linearized} 
\end{figure}

Lastly we consider the high rotation speed torque data and its comparison with the prediction of Eq.~(\ref{eq:Torque}).  For this, we simply subtract the quasistatic torque and plot the result $\tau-\tau_0$ as a function of $x=(1/32) \rho_g D_c {L_c}^4\omega^2$ in Figs.~5a,b for the two different grain types.  According to Eq.~(\ref{eq:Torque}), both axes have units of torque and {\it all} data should collapse to the line $\tau-\tau_0=\beta x$.  Indeed this is consistent with observation, showing that Eq.~(\ref{eq:Torque}) correctly predicts the materials and geometry dependence of torque in the inertial regime.  Furthermore, the fits to a line are consistent with a single value $\beta=0.6\pm0.2$ for both grain types.  This is somewhat smaller than the result $\beta=1.0\pm0.1$ of Ref.~\cite{Katsuragi:2007p181} based on impact, perhaps for one of the reasons noted above.  Note, however, that since $\alpha$ and $\beta$ are {\it both} smaller for rotating rods than for downward spheres, it seems fair to conclude that the drag on the latter is larger.

\section{Conclusion}

To summarize, we have developed a new method to study granular drag. Our system achieves a reproducible steady-state, facilitating statistically robust data collection, and enables us to vary variables such as speed and depth independently and in situ. This system has the potential to be a powerful tool in understanding the fundamental bulk mechanics of granular flow: already our results have shed light on the grain-scale mechanics which dictate drag forces on the scale of an intruder. We observe drag forces proportional to the rate squared for high rotation rates with magnitudes which depend only upon the geometry of the intruder and the density of the granular packing, suggesting drag forces at high speeds are inertial in origin. In the slow quasistatic limit we confirm that the drag force is rate-independent, and scales like a hydrostatic pressure. This scaling suggests that the response of a granular packing to low-speed intruders is set by the strength of static contact forces. We modify the loading of contacts between grains in the packing with an applied pressure drop across the bed, and the agreement between data and Eq.~(\ref{eq:pressure}) provides further corroboration for our hypothesis. A simple model based on this interpretation of the mechanistic origins of granular drag has proven effective in describing impact dynamics~\cite{Katsuragi:2007p181}. That we have similar success in applying the model to describe the drag experienced by an intruder moving horizontally at a constant rate reinforces the supposition that the nature of granular drag is universal, independent of the details of the system.

An obvious open question is why the values we obtain for $\alpha$ and $\beta$ are different from those reported in~\cite{Katsuragi:2007p181}. This disparity suggests that Eq.~(\ref{eq:Torque}) might not capture the full geometry dependence of the drag force.  As discussed above, one such geometric effect might be the dependence of drag forces on the direction of intruder motion relative to the direction of gravity. Another aspect of geometry that might warrant closer examination is intruder shape. Cross-sectional area may not be sufficient to fully describe the role of intruder shape, and corrections for three-dimensional features of intruders may prove important. A further characterization of granular drag forces in a more extensive set of geometries is needed to fully determine the role of such geometric effects.

Another area of research that remains largely unexplored is the full form of the drag force in the fluidized case. While gas-fluidization is a fairly specific phenomena, it's also a useful way to induce flow, and understanding the non-inertial components of the drag force in a gas fluidized bed may provide insight into the behaviors of other granular systems in which there are no lasting static contacts and grain-grain interactions tend to be ballistic, such as in very rapid and low-density granular flows. As such, an in depth investigation of low-rate drag forces for $U>U_c$ would be of great interest.

Lastly, understanding the effect of more complicated grain interactions upon the drag dynamics in a bed would be essential to describe drag in many real-world granular systems. Our model assumes that the only mechanisms that contribute to drag are inertial interactions between the intruder and the grains, and static contact forces between grains. For our system - mechanically simple, hard, dry, spherical grains - these are good assumptions, but many granular systems are not nearly so simple. A similar approach to characterizing drag in systems of soft grains, or in which a more viscous interstitial fluid is introduced would provide insight that might prove invaluable in application.

\begin{acknowledgments}
We thank Hiroaki Katsuragi for helpful discussion and advice.  Our work was supported by the National Science Foundation through grant DMR-0704147.
\end{acknowledgments}


\bibliography{Manuscript}

\begin{thebibliography}{34}
\expandafter\ifx\csname natexlab\endcsname\relax\def\natexlab#1{#1}\fi
\expandafter\ifx\csname bibnamefont\endcsname\relax
  \def\bibnamefont#1{#1}\fi
\expandafter\ifx\csname bibfnamefont\endcsname\relax
  \def\bibfnamefont#1{#1}\fi
\expandafter\ifx\csname citenamefont\endcsname\relax
  \def\citenamefont#1{#1}\fi
\expandafter\ifx\csname url\endcsname\relax
  \def\url#1{\texttt{#1}}\fi
\expandafter\ifx\csname urlprefix\endcsname\relax\def\urlprefix{URL }\fi
\providecommand{\bibinfo}[2]{#2}
\providecommand{\eprint}[2][]{\url{#2}}

\bibitem[{\citenamefont{Wieghardt}(1975)}]{Wieghardt:1975p7003}
\bibinfo{author}{\bibfnamefont{K.}~\bibnamefont{Wieghardt}},
  \bibinfo{journal}{Annu. Rev. Fluid Mech.} pp. \bibinfo{pages}{89--113}
  (\bibinfo{year}{1975}).

\bibitem[{\citenamefont{Thompson and Grest}(1991)}]{Thompson:1991p6993}
\bibinfo{author}{\bibfnamefont{P.~A.} \bibnamefont{Thompson}} \bibnamefont{and}
  \bibinfo{author}{\bibfnamefont{G.~S.} \bibnamefont{Grest}},
  \bibinfo{journal}{Phys. Rev. Lett.} \textbf{\bibinfo{volume}{67}},
  \bibinfo{pages}{1751} (\bibinfo{year}{1991}).

\bibitem[{\citenamefont{Fenistein and van Hecke}(2003)}]{Fenistein:2003p6914}
\bibinfo{author}{\bibfnamefont{D.}~\bibnamefont{Fenistein}} \bibnamefont{and}
  \bibinfo{author}{\bibfnamefont{M.}~\bibnamefont{van Hecke}},
  \bibinfo{journal}{Nature} \textbf{\bibinfo{volume}{425}},
  \bibinfo{pages}{256} (\bibinfo{year}{2003}).

\bibitem[{\citenamefont{Behringer et~al.}(2001)\citenamefont{Behringer,
  Clement, Geng, Howell, Kondic, Metcalfe, O'Hern, Reydellet, Tennakoon, Vanel
  et~al.}}]{Behringer:2001p7187}
\bibinfo{author}{\bibfnamefont{R.~P.} \bibnamefont{Behringer}},
  \bibinfo{author}{\bibfnamefont{E.}~\bibnamefont{Clement}},
  \bibinfo{author}{\bibfnamefont{J.}~\bibnamefont{Geng}},
  \bibinfo{author}{\bibfnamefont{D.}~\bibnamefont{Howell}},
  \bibinfo{author}{\bibfnamefont{L.}~\bibnamefont{Kondic}},
  \bibinfo{author}{\bibfnamefont{G.}~\bibnamefont{Metcalfe}},
  \bibinfo{author}{\bibfnamefont{C.}~\bibnamefont{O'Hern}},
  \bibinfo{author}{\bibfnamefont{G.}~\bibnamefont{Reydellet}},
  \bibinfo{author}{\bibfnamefont{S.}~\bibnamefont{Tennakoon}},
  \bibinfo{author}{\bibfnamefont{L.}~\bibnamefont{Vanel}},
  \bibnamefont{et~al.}, \emph{\bibinfo{title}{Science in the Sandbox:
  Fluctuations, Friction and Instabilities}}, vol. \bibinfo{volume}{567} of
  \emph{\bibinfo{series}{Lecture Notes in Physics}}
  (\bibinfo{publisher}{Springer-Verlag}, \bibinfo{address}{Berlin},
  \bibinfo{year}{2001}).

\bibitem[{\citenamefont{Liu and Nagel}(1998)}]{Liu:1998p7104}
\bibinfo{author}{\bibfnamefont{A.~J.} \bibnamefont{Liu}} \bibnamefont{and}
  \bibinfo{author}{\bibfnamefont{S.~R.} \bibnamefont{Nagel}},
  \bibinfo{journal}{Nature} \textbf{\bibinfo{volume}{396}}, \bibinfo{pages}{21}
  (\bibinfo{year}{1998}).

\bibitem[{\citenamefont{Boudet et~al.}(2006)\citenamefont{Boudet, Amarouchene,
  and Kellay}}]{Boudet:2006p7070}
\bibinfo{author}{\bibfnamefont{J.~F.} \bibnamefont{Boudet}},
  \bibinfo{author}{\bibfnamefont{Y.}~\bibnamefont{Amarouchene}},
  \bibnamefont{and} \bibinfo{author}{\bibfnamefont{H.}~\bibnamefont{Kellay}},
  \bibinfo{journal}{Phys. Rev. Lett.} \textbf{\bibinfo{volume}{96}},
  \bibinfo{pages}{158001} (\bibinfo{year}{2006}).

\bibitem[{\citenamefont{Walsh et~al.}(2003{\natexlab{a}})\citenamefont{Walsh,
  Holloway, Habdas, and de~Bruyn}}]{Walsh:2003p7129}
\bibinfo{author}{\bibfnamefont{A.~M.} \bibnamefont{Walsh}},
  \bibinfo{author}{\bibfnamefont{K.~E.} \bibnamefont{Holloway}},
  \bibinfo{author}{\bibfnamefont{P.}~\bibnamefont{Habdas}}, \bibnamefont{and}
  \bibinfo{author}{\bibfnamefont{J.~R.} \bibnamefont{de~Bruyn}},
  \bibinfo{journal}{Phys. Rev. Lett.} \textbf{\bibinfo{volume}{91}},
  \bibinfo{pages}{104301} (\bibinfo{year}{2003}{\natexlab{a}}).

\bibitem[{\citenamefont{Amato and Williams}(1998)}]{Amato:1998p7077}
\bibinfo{author}{\bibfnamefont{J.~C.} \bibnamefont{Amato}} \bibnamefont{and}
  \bibinfo{author}{\bibfnamefont{R.~E.} \bibnamefont{Williams}},
  \bibinfo{journal}{Am. J. Phys.} \textbf{\bibinfo{volume}{66}},
  \bibinfo{pages}{141} (\bibinfo{year}{1998}).

\bibitem[{\citenamefont{Daniels et~al.}(2004)\citenamefont{Daniels, Coppock,
  and Behringer}}]{Daniels:2004p7074}
\bibinfo{author}{\bibfnamefont{K.~E.} \bibnamefont{Daniels}},
  \bibinfo{author}{\bibfnamefont{J.~E.} \bibnamefont{Coppock}},
  \bibnamefont{and} \bibinfo{author}{\bibfnamefont{R.~P.}
  \bibnamefont{Behringer}}, \bibinfo{journal}{Chaos}
  \textbf{\bibinfo{volume}{14}} (\bibinfo{year}{2004}).

\bibitem[{\citenamefont{Zheng et~al.}(2004)\citenamefont{Zheng, Wang, and
  Qiu}}]{Zheng:2004p7302}
\bibinfo{author}{\bibfnamefont{X.-J.} \bibnamefont{Zheng}},
  \bibinfo{author}{\bibfnamefont{Z.-T.} \bibnamefont{Wang}}, \bibnamefont{and}
  \bibinfo{author}{\bibfnamefont{Z.-G.} \bibnamefont{Qiu}},
  \bibinfo{journal}{Euro. Phys. J. E} \textbf{\bibinfo{volume}{13}},
  \bibinfo{pages}{321} (\bibinfo{year}{2004}).

\bibitem[{\citenamefont{Goldman and Umbanhowar}(2008)}]{Goldman:2008p7403}
\bibinfo{author}{\bibfnamefont{D.~I.} \bibnamefont{Goldman}} \bibnamefont{and}
  \bibinfo{author}{\bibfnamefont{P.}~\bibnamefont{Umbanhowar}},
  \bibinfo{journal}{Phys. Rev. E} \textbf{\bibinfo{volume}{77}},
  \bibinfo{pages}{021308} (\bibinfo{year}{2008}).

\bibitem[{\citenamefont{Uehara et~al.}(2003)\citenamefont{Uehara, Ambroso,
  Ojha, and Durian}}]{Uehara:2003p7073}
\bibinfo{author}{\bibfnamefont{J.~S.} \bibnamefont{Uehara}},
  \bibinfo{author}{\bibfnamefont{M.~A.} \bibnamefont{Ambroso}},
  \bibinfo{author}{\bibfnamefont{R.~P.} \bibnamefont{Ojha}}, \bibnamefont{and}
  \bibinfo{author}{\bibfnamefont{D.~J.} \bibnamefont{Durian}},
  \bibinfo{journal}{Phys. Rev. Lett.} \textbf{\bibinfo{volume}{90}},
  \bibinfo{pages}{194301} (\bibinfo{year}{2003}).

\bibitem[{\citenamefont{Walsh and de~Bruyn}(2004)}]{Walsh:2004p7304}
\bibinfo{author}{\bibfnamefont{A.~M.} \bibnamefont{Walsh}} \bibnamefont{and}
  \bibinfo{author}{\bibfnamefont{J.~R.} \bibnamefont{de~Bruyn}},
  \bibinfo{journal}{Can. J. Phys.} \textbf{\bibinfo{volume}{82}},
  \bibinfo{pages}{439} (\bibinfo{year}{2004}).

\bibitem[{\citenamefont{Ciamarra et~al.}(2004)\citenamefont{Ciamarra, Lara,
  Lee, Goldman, Vishik, and Swinney}}]{PicaCiamarra:2004p7335}
\bibinfo{author}{\bibfnamefont{M.~P.} \bibnamefont{Ciamarra}},
  \bibinfo{author}{\bibfnamefont{A.~H.} \bibnamefont{Lara}},
  \bibinfo{author}{\bibfnamefont{A.~T.} \bibnamefont{Lee}},
  \bibinfo{author}{\bibfnamefont{D.~I.} \bibnamefont{Goldman}},
  \bibinfo{author}{\bibfnamefont{I.}~\bibnamefont{Vishik}}, \bibnamefont{and}
  \bibinfo{author}{\bibfnamefont{H.~L.} \bibnamefont{Swinney}},
  \bibinfo{journal}{Phys. Rev. Lett.} \textbf{\bibinfo{volume}{92}},
  \bibinfo{pages}{194301} (\bibinfo{year}{2004}).

\bibitem[{\citenamefont{Lohse et~al.}(2004{\natexlab{a}})\citenamefont{Lohse,
  Rauh{\'e}, Bergmann, and van~der Meer}}]{Lohse:2004p7357}
\bibinfo{author}{\bibfnamefont{D.}~\bibnamefont{Lohse}},
  \bibinfo{author}{\bibfnamefont{R.}~\bibnamefont{Rauh{\'e}}},
  \bibinfo{author}{\bibfnamefont{R.}~\bibnamefont{Bergmann}}, \bibnamefont{and}
  \bibinfo{author}{\bibfnamefont{D.}~\bibnamefont{van~der Meer}},
  \bibinfo{journal}{Nature} \textbf{\bibinfo{volume}{432}},
  \bibinfo{pages}{689} (\bibinfo{year}{2004}{\natexlab{a}}).

\bibitem[{\citenamefont{Hou et~al.}(2005)\citenamefont{Hou, Peng, Liu, Lu, and
  Chan}}]{Hou:2005p7358}
\bibinfo{author}{\bibfnamefont{M.}~\bibnamefont{Hou}},
  \bibinfo{author}{\bibfnamefont{Z.}~\bibnamefont{Peng}},
  \bibinfo{author}{\bibfnamefont{R.}~\bibnamefont{Liu}},
  \bibinfo{author}{\bibfnamefont{K.}~\bibnamefont{Lu}}, \bibnamefont{and}
  \bibinfo{author}{\bibfnamefont{C.~K.} \bibnamefont{Chan}},
  \bibinfo{journal}{Phys. Rev. E} \textbf{\bibinfo{volume}{72}},
  \bibinfo{pages}{062301} (\bibinfo{year}{2005}).

\bibitem[{\citenamefont{Walsh et~al.}(2003{\natexlab{b}})\citenamefont{Walsh,
  Holloway, Habdas, and de~Bruyn}}]{Walsh:2003p7076}
\bibinfo{author}{\bibfnamefont{A.~M.} \bibnamefont{Walsh}},
  \bibinfo{author}{\bibfnamefont{K.~E.} \bibnamefont{Holloway}},
  \bibinfo{author}{\bibfnamefont{P.}~\bibnamefont{Habdas}}, \bibnamefont{and}
  \bibinfo{author}{\bibfnamefont{J.~R.} \bibnamefont{de~Bruyn}},
  \bibinfo{journal}{Phys. Rev. Lett.} \textbf{\bibinfo{volume}{91}},
  \bibinfo{pages}{104301} (\bibinfo{year}{2003}{\natexlab{b}}).

\bibitem[{\citenamefont{Katsuragi and Durian}(2007)}]{Katsuragi:2007p181}
\bibinfo{author}{\bibfnamefont{H.}~\bibnamefont{Katsuragi}} \bibnamefont{and}
  \bibinfo{author}{\bibfnamefont{D.~J.} \bibnamefont{Durian}},
  \bibinfo{journal}{Nature Phys.} \textbf{\bibinfo{volume}{3}},
  \bibinfo{pages}{420 } (\bibinfo{year}{2007}).

\bibitem[{\citenamefont{Albert et~al.}(1999)\citenamefont{Albert, Pfeifer,
  Barab{\'a}si, and Schiffer}}]{Albert:1999p623}
\bibinfo{author}{\bibfnamefont{R.}~\bibnamefont{Albert}},
  \bibinfo{author}{\bibfnamefont{M.~A.} \bibnamefont{Pfeifer}},
  \bibinfo{author}{\bibfnamefont{A.-L.} \bibnamefont{Barab{\'a}si}},
  \bibnamefont{and} \bibinfo{author}{\bibfnamefont{P.}~\bibnamefont{Schiffer}},
  \bibinfo{journal}{Phys. Rev. Lett.} \textbf{\bibinfo{volume}{82}},
  \bibinfo{pages}{205} (\bibinfo{year}{1999}).

\bibitem[{\citenamefont{Albert et~al.}(2000)\citenamefont{Albert, Tegzes,
  Kahng, Albert, Sample, Pfeifer, Barabasi, Vicsek, and
  Schiffer}}]{Albert:2000p620}
\bibinfo{author}{\bibfnamefont{I.}~\bibnamefont{Albert}},
  \bibinfo{author}{\bibfnamefont{P.}~\bibnamefont{Tegzes}},
  \bibinfo{author}{\bibfnamefont{B.}~\bibnamefont{Kahng}},
  \bibinfo{author}{\bibfnamefont{R.}~\bibnamefont{Albert}},
  \bibinfo{author}{\bibfnamefont{J.~G.} \bibnamefont{Sample}},
  \bibinfo{author}{\bibfnamefont{M.}~\bibnamefont{Pfeifer}},
  \bibinfo{author}{\bibfnamefont{A.-L.} \bibnamefont{Barabasi}},
  \bibinfo{author}{\bibfnamefont{T.}~\bibnamefont{Vicsek}}, \bibnamefont{and}
  \bibinfo{author}{\bibfnamefont{P.}~\bibnamefont{Schiffer}},
  \bibinfo{journal}{Phys. Rev. Lett.} \textbf{\bibinfo{volume}{84}},
  \bibinfo{pages}{5122} (\bibinfo{year}{2000}).

\bibitem[{\citenamefont{Albert et~al.}(2001{\natexlab{a}})\citenamefont{Albert,
  Tegzes, Albert, Sample, Barabasi, Vicsek, Kahng, and
  Schiffer}}]{Albert:2001p1481}
\bibinfo{author}{\bibfnamefont{I.}~\bibnamefont{Albert}},
  \bibinfo{author}{\bibfnamefont{P.}~\bibnamefont{Tegzes}},
  \bibinfo{author}{\bibfnamefont{R.}~\bibnamefont{Albert}},
  \bibinfo{author}{\bibfnamefont{J.~G.} \bibnamefont{Sample}},
  \bibinfo{author}{\bibfnamefont{A.-L.} \bibnamefont{Barabasi}},
  \bibinfo{author}{\bibfnamefont{T.}~\bibnamefont{Vicsek}},
  \bibinfo{author}{\bibfnamefont{B.}~\bibnamefont{Kahng}}, \bibnamefont{and}
  \bibinfo{author}{\bibfnamefont{P.}~\bibnamefont{Schiffer}},
  \bibinfo{journal}{Phys. Rev. E} \textbf{\bibinfo{volume}{64}},
  \bibinfo{pages}{031307} (\bibinfo{year}{2001}{\natexlab{a}}).

\bibitem[{\citenamefont{Albert et~al.}(2001{\natexlab{b}})\citenamefont{Albert,
  Sample, Morss, Rajagopalan, Barabasi, and Schiffer}}]{Albert:2001p2}
\bibinfo{author}{\bibfnamefont{I.}~\bibnamefont{Albert}},
  \bibinfo{author}{\bibfnamefont{J.~G.} \bibnamefont{Sample}},
  \bibinfo{author}{\bibfnamefont{A.~J.} \bibnamefont{Morss}},
  \bibinfo{author}{\bibfnamefont{S.}~\bibnamefont{Rajagopalan}},
  \bibinfo{author}{\bibfnamefont{A.-L.} \bibnamefont{Barabasi}},
  \bibnamefont{and} \bibinfo{author}{\bibfnamefont{P.}~\bibnamefont{Schiffer}},
  \bibinfo{journal}{Phys. Rev. E} \textbf{\bibinfo{volume}{64}},
  \bibinfo{pages}{061303} (\bibinfo{year}{2001}{\natexlab{b}}).

\bibitem[{\citenamefont{Remy et~al.}(2009)\citenamefont{Remy, Khinast, and
  Glasser}}]{Remy:2009p8064}
\bibinfo{author}{\bibfnamefont{B.}~\bibnamefont{Remy}},
  \bibinfo{author}{\bibfnamefont{J.~G.} \bibnamefont{Khinast}},
  \bibnamefont{and} \bibinfo{author}{\bibfnamefont{B.~J.}
  \bibnamefont{Glasser}}, \bibinfo{journal}{Am. Inst. Chem. Eng.}
  \textbf{\bibinfo{volume}{55}}, \bibinfo{pages}{2035} (\bibinfo{year}{2009}).

\bibitem[{\citenamefont{Jaworski et~al.}(2001)\citenamefont{Jaworski, Dyster,
  and Nienow}}]{Jaworski:2001p8338}
\bibinfo{author}{\bibfnamefont{Z.}~\bibnamefont{Jaworski}},
  \bibinfo{author}{\bibfnamefont{K.~N.} \bibnamefont{Dyster}},
  \bibnamefont{and} \bibinfo{author}{\bibfnamefont{A.~W.}
  \bibnamefont{Nienow}}, \bibinfo{journal}{Trans IChemE}
  \textbf{\bibinfo{volume}{79}}, \bibinfo{pages}{887} (\bibinfo{year}{2001}).

\bibitem[{\citenamefont{Stewart et~al.}(2001)\citenamefont{Stewart,
  Bridgewater, and Parker}}]{Stewart:2001p4257}
\bibinfo{author}{\bibfnamefont{R.~L.} \bibnamefont{Stewart}},
  \bibinfo{author}{\bibfnamefont{J.}~\bibnamefont{Bridgewater}},
  \bibnamefont{and} \bibinfo{author}{\bibfnamefont{D.~J.}
  \bibnamefont{Parker}}, \bibinfo{journal}{Chem. Eng. Science}
  \textbf{\bibinfo{volume}{56}}, \bibinfo{pages}{4257} (\bibinfo{year}{2001}).

\bibitem[{\citenamefont{Bagster and Bridgewater}(1967)}]{Bagster:1967p1899}
\bibinfo{author}{\bibfnamefont{D.~F.} \bibnamefont{Bagster}} \bibnamefont{and}
  \bibinfo{author}{\bibfnamefont{J.}~\bibnamefont{Bridgewater}},
  \bibinfo{journal}{Powder Tech.} \textbf{\bibinfo{volume}{1}},
  \bibinfo{pages}{189} (\bibinfo{year}{1967}).

\bibitem[{\citenamefont{Barnes and Nguyen}(2001)}]{Barnes:2001p9297}
\bibinfo{author}{\bibfnamefont{H.~A.} \bibnamefont{Barnes}} \bibnamefont{and}
  \bibinfo{author}{\bibfnamefont{Q.~D.} \bibnamefont{Nguyen}},
  \bibinfo{journal}{J. Non-Newtonian Fluid Mech.}
  \textbf{\bibinfo{volume}{98}}, \bibinfo{pages}{1} (\bibinfo{year}{2001}).

\bibitem[{\citenamefont{Daniel et~al.}(2008)\citenamefont{Daniel, Poloski, and
  Saez}}]{Daniel:2008p9330}
\bibinfo{author}{\bibfnamefont{R.~C.} \bibnamefont{Daniel}},
  \bibinfo{author}{\bibfnamefont{A.~P.} \bibnamefont{Poloski}},
  \bibnamefont{and} \bibinfo{author}{\bibfnamefont{A.~E.} \bibnamefont{Saez}},
  \bibinfo{journal}{Powder Tech.} \textbf{\bibinfo{volume}{181}},
  \bibinfo{pages}{237} (\bibinfo{year}{2008}).

\bibitem[{\citenamefont{Soller and Koehler}(2006)}]{Soller:2006p8632}
\bibinfo{author}{\bibfnamefont{R.}~\bibnamefont{Soller}} \bibnamefont{and}
  \bibinfo{author}{\bibfnamefont{S.~A.} \bibnamefont{Koehler}},
  \bibinfo{journal}{Phys. Rev. E} \textbf{\bibinfo{volume}{74}},
  \bibinfo{pages}{021305} (\bibinfo{year}{2006}).

\bibitem[{\citenamefont{Poloski et~al.}(2006)\citenamefont{Poloski, Bredt,
  Daniel, and Saez}}]{Poloski:2006p9438}
\bibinfo{author}{\bibfnamefont{A.~P.} \bibnamefont{Poloski}},
  \bibinfo{author}{\bibfnamefont{P.~R.} \bibnamefont{Bredt}},
  \bibinfo{author}{\bibfnamefont{R.~C.} \bibnamefont{Daniel}},
  \bibnamefont{and} \bibinfo{author}{\bibfnamefont{A.~E.} \bibnamefont{Saez}},
  \bibinfo{journal}{Rheo. Acta} \textbf{\bibinfo{volume}{46}},
  \bibinfo{pages}{249} (\bibinfo{year}{2006}).

\bibitem[{\citenamefont{Pak et~al.}(1995)\citenamefont{Pak, Doorn, and
  Behringer}}]{Pak:1995}
\bibinfo{author}{\bibfnamefont{H.~K.} \bibnamefont{Pak}},
  \bibinfo{author}{\bibfnamefont{E.~V.} \bibnamefont{Doorn}}, \bibnamefont{and}
  \bibinfo{author}{\bibfnamefont{R.~P.} \bibnamefont{Behringer}},
  \bibinfo{journal}{Phys. Rev. Lett.} \textbf{\bibinfo{volume}{74}},
  \bibinfo{pages}{4643} (\bibinfo{year}{1995}).

\bibitem[{\citenamefont{Royer et~al.}(2005)\citenamefont{Royer, Corwin, Flior,
  Cordero, Rivers, Eng, and Jaeger}}]{Royer:2005}
\bibinfo{author}{\bibfnamefont{J.~R.} \bibnamefont{Royer}},
  \bibinfo{author}{\bibfnamefont{E.~I.} \bibnamefont{Corwin}},
  \bibinfo{author}{\bibfnamefont{A.}~\bibnamefont{Flior}},
  \bibinfo{author}{\bibfnamefont{M.}~\bibnamefont{Cordero}},
  \bibinfo{author}{\bibfnamefont{M.~L.} \bibnamefont{Rivers}},
  \bibinfo{author}{\bibfnamefont{P.~J.} \bibnamefont{Eng}}, \bibnamefont{and}
  \bibinfo{author}{\bibfnamefont{H.~M.} \bibnamefont{Jaeger}},
  \bibinfo{journal}{Nature Physics} \textbf{\bibinfo{volume}{1}},
  \bibinfo{pages}{164} (\bibinfo{year}{2005}).

\bibitem[{\citenamefont{Lohse et~al.}(2004{\natexlab{b}})\citenamefont{Lohse,
  Bergmann, Mikkelsen, Zeilstra, van~der Meer, Versluis, van~der Weele, van~der
  Hoef, and Kuipers}}]{Lohse:1234}
\bibinfo{author}{\bibfnamefont{D.}~\bibnamefont{Lohse}},
  \bibinfo{author}{\bibfnamefont{R.}~\bibnamefont{Bergmann}},
  \bibinfo{author}{\bibfnamefont{R.}~\bibnamefont{Mikkelsen}},
  \bibinfo{author}{\bibfnamefont{C.}~\bibnamefont{Zeilstra}},
  \bibinfo{author}{\bibfnamefont{D.}~\bibnamefont{van~der Meer}},
  \bibinfo{author}{\bibfnamefont{M.}~\bibnamefont{Versluis}},
  \bibinfo{author}{\bibfnamefont{K.}~\bibnamefont{van~der Weele}},
  \bibinfo{author}{\bibfnamefont{M.}~\bibnamefont{van~der Hoef}},
  \bibnamefont{and} \bibinfo{author}{\bibfnamefont{H.}~\bibnamefont{Kuipers}},
  \bibinfo{journal}{Phys. Rev. Lett.} \textbf{\bibinfo{volume}{93}},
  \bibinfo{pages}{198003} (\bibinfo{year}{2004}{\natexlab{b}}).

\bibitem[{\citenamefont{Ojha et~al.}(2000)\citenamefont{Ojha, Menon, and
  Durian}}]{Ojha:2000p94}
\bibinfo{author}{\bibfnamefont{R.}~\bibnamefont{Ojha}},
  \bibinfo{author}{\bibfnamefont{N.}~\bibnamefont{Menon}}, \bibnamefont{and}
  \bibinfo{author}{\bibfnamefont{D.~J.} \bibnamefont{Durian}},
  \bibinfo{journal}{Phys. Rev. E} \textbf{\bibinfo{volume}{62}},
  \bibinfo{pages}{4442} (\bibinfo{year}{2000}).

\end{thebibliography}

\end{document}